\documentclass[number,sort&compress]{elsarticle}

\usepackage{hyperref}
\usepackage{graphicx}
\DeclareGraphicsExtensions{.pdf,.png,.jpg} 
\usepackage{fancyvrb}
\usepackage{amsmath}
\usepackage{placeins}
\usepackage{float}
\usepackage{url}

\begin{document}

\begin{frontmatter}

\title{Development of an isotropic optical light source for testing nuclear instruments}

\cortext[mycorrespondingauthor]{Corresponding author}
\author{Z. W. Yokley\corref{mycorrespondingauthor}}
\ead{zwyokley@vt.edu}
\author{S. D. Rountree\corref{cor2}}
\author{R. B. Vogelaar\corref{cor2}}
\address{Center for Neutrino Physics, Virginia Tech, Blacksburg, VA 24061, USA}


\begin{abstract}
	Nuclear instruments that require precise characterization and calibration of their optical components need well-characterized optical light sources with the desired wavelength, intensity, and directivity. This paper presents a novel technique for determining the performance of optical components by producing an isotropic-like source with a robotically positioned LED. The theory of operation for this light source, results of Monte Carlo validation studies, and experimental results are presented.
\end{abstract}


\end{frontmatter}

\section{Introduction}
\label{sec:Intro}

Optical components such as photomultiplier tubes (PMTs) and light guides are a common part of nuclear instrumentation. Often these components need precise calibration and characterization to ensure adequate operation, thus they require calibration light sources with appropriate wavelength, intensity, and a well-known angular emission distribution. Common solutions are lasers or LEDs which are often coupled to optical fiber or diffusers \cite{Ritter:2010zz, Moffat:2005tq}. However, these light sources are typically a few centimeters in diameter and made for \textit{in situ} calibrations of ton and kiloton-scale detectors where they approximate isotropic point sources. For testing individual components, however, these light sources are no longer isotropic point sources because their dimensions are usually of the same order as the components. While using smaller diffusers is one possible solution to the problem, they can still be insufficient for producing isotropic light. In order to move past this problem a new concept for an effective isotropic point source of light was developed using a robotically positioned LED. 
	
This paper is organized as follows: Section \ref{sec:2_Concept} discusses the concept behind the robotic light source; Section \ref{sec:3_MC} presents the results of a Monte Carlo (MC) validation study; Section \ref{sec:4_Setup} presents the experimental setup with the as-built light source; and Section \ref{sec:5_Results} presents the results from the source; Section \ref{sec:6_Conclusions} gives some concluding remarks concerning the approach presented here and future work on the use of robotic light sources.
	
\section{Concept}
\label{sec:2_Concept}

The robotic light source was inspired by the modeling of an isotropic point source in a MC simulation. To model such a source, two angles are used to define the propagation direction for each emitted photon: the zenith (or polar) angle $ \theta $ and the azimuthal angle $ \phi $. These angles range from $ \left[0,\pi\right]  $ and $ \left[0,2\pi\right)  $ respectively. The angles are sampled by picking $ \phi $ from a uniform distribution on $ \left[0,2\pi\right)  $ and $ \cos\theta $ from a uniform distribution on $ \left[-1,1\right] $. Now consider an optical light source that emits light in only one direction; a time integral isotropic source is produced by positioning it like the MC model just discussed. Equivalently one can point the source in all directions for an equal time duration to produce an effective isotropic source. This method also works with any real anisotropic source. 
	
For concreteness, consider an LED that emits light into a finite solid angle. Since the light is emitted into a finite solid angle, then the LED can be pointed in a limited number of directions to approximate the probability density functions (PDFs) for $ \phi $ and $ \theta $. This choice is equivalent to discretizing the PDFs. For an accurate representation of the PDF's, the size of the discretization should be of the same size or smaller than the solid angle that the LED emits into. This ensures adequate overlap between the bins. Furthermore, if the directions are not equally probable and they are each run for equal time durations, then the detector responses require statistical weights that are based on the PDFs. Specifically, the weights are proportional the integrals of each discrete bin of the PDFs. 	
	
\begin{figure}[h] 
		\centering
		\includegraphics[scale=0.35, angle=-90.0]{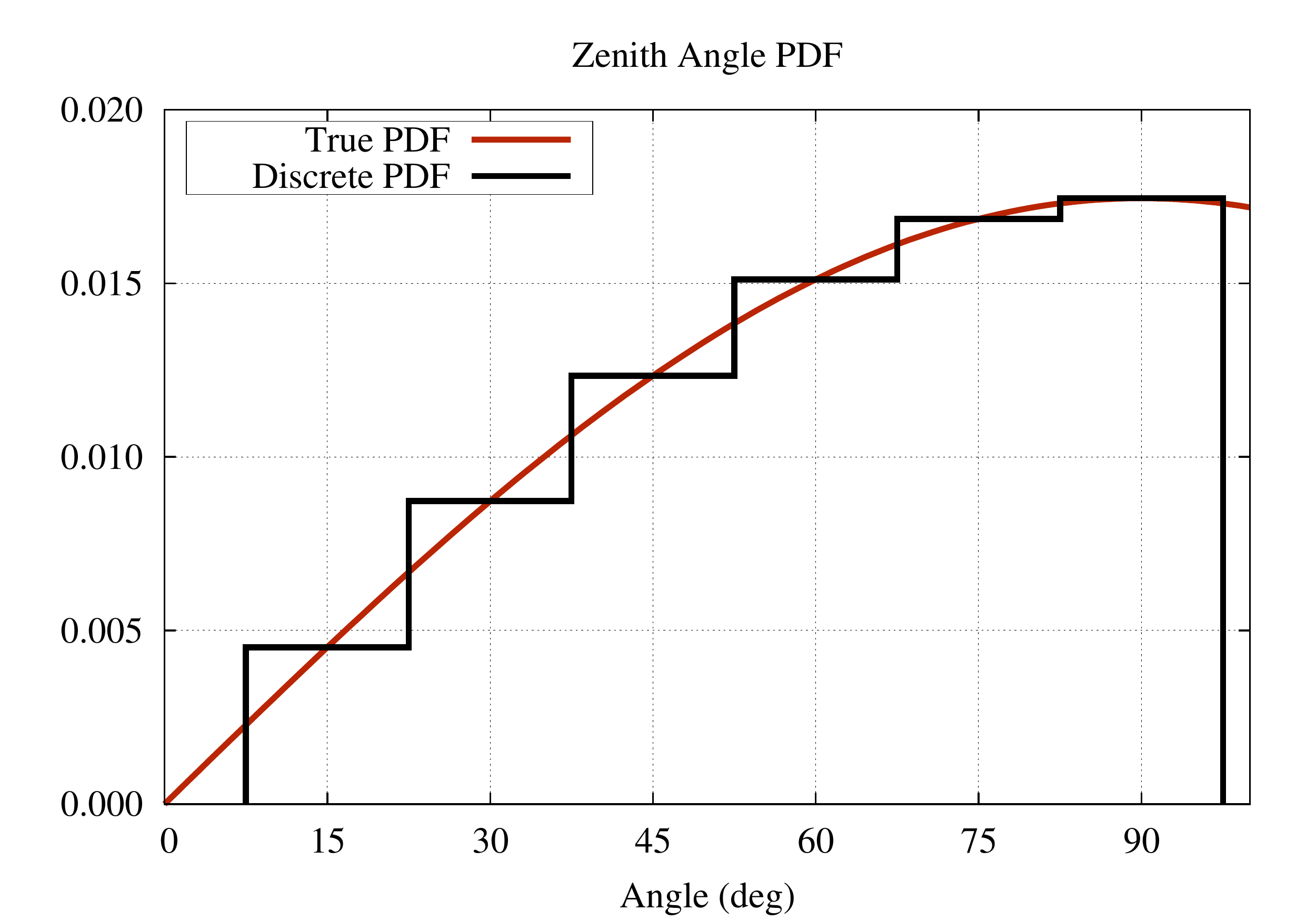} 
		\caption{A plot of the zenith angle PDF and the discrete approximation for the chosen directions. The directions define the central angles of each bin.} 
		\label{fig:discrete}
	\end{figure}

For an example, the following were chosen: $ \theta = 15,\,30,\, 45,\, 60,\, 75,\, $ and $ 90^{\circ} $ and $ \phi $ uniform on $ \left[0,2\pi\right)  $; therefore, only the $ \theta $ PDF is discretized. A plot of the $ \theta $ PDF and its discrete approximation is shown in Figure \ref{fig:discrete}.
	\FloatBarrier	
The choice of these directions was made for a Lambertian emitter, which emits light into a fairly significant solid angle. From Figure \ref{fig:discrete} it is clear that the chosen directions are not equally probable; therefore, their statistical weights need to be proportional to the value of the zenith angle's PDF evaluated at each $ \theta $:
	\begin{equation}
		\label{eqn:weights0}
		w_{15^{\circ}} = \sin\left( 15^{\circ}\right),\; w_{30^{\circ}} = \sin\left( 30^{\circ}\right), \;\mathrm{etc.}
	\end{equation}
Since the proportionality holds when the wights are multiplied by a constant, then, for convenience, take
	\begin{equation}
		\label{eqn:weights1}
		w_{i} \mapsto \frac{w_{i}}{\sum_{i}w_{i}}.
\end{equation}
From Equation \ref{eqn:weights1}, it is clear that a weighted average of detector responses is performed for each direction. The weights defined by equations \ref{eqn:weights0} and \ref{eqn:weights1} are shown in Table \ref{tab:statWeights}.
		\FloatBarrier
		\begin{table}[htbp]
			\centering
			\caption{Statistical weights for listed directions: $ \theta = 15,\,30,\, 45,\, 60,\, 75,\, $ and $ 90^{\circ} $ and $ \phi $ uniform on $ \left[0,2\pi\right)  $.}
			\label{tab:statWeights}			
			\begin{tabular}{ll}
				\hline \hline
				Zenith Angle ($ \deg $)	&	Weight		\\ \hline
				15						&	0.0602		\\
				30						&	0.1163		\\
				45						&	0.1645		\\
				60						&	0.2015		\\
				75						&	0.2247		\\
				90						&	0.2327	
				\\ \hline \hline
			\end{tabular}
		\end{table}
		\FloatBarrier
The advantage of using this (or a similar method) is that an effective isotropic light source can be made more quickly than randomly pointing and that analysis become simpler. To see this, consider that for each direction an average number of photons will be detected. These averages enter into the weighted average just discussed, and the result behaves as if the light source was isotropic. This result is defined as the \textit{proxy number detected}. An additional advantage of this method is that it can be applied to any PDF, thus it is useful for making an approximation of any light source.

\section{Monte Carlo validation}
\label{sec:3_MC}

\begin{figure}[h] 
				\centering
		\includegraphics[scale=0.5, angle=0.0]{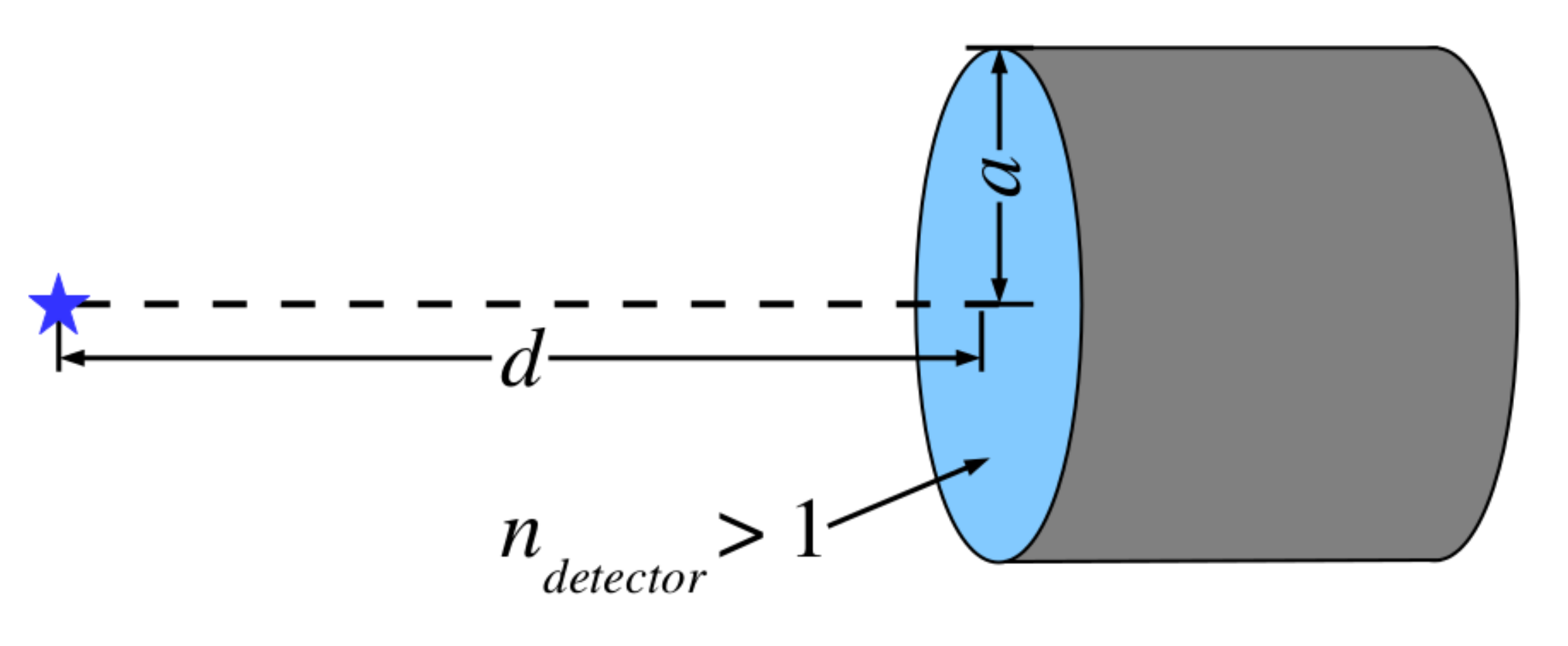} 
		\caption{The geometry of the light source (blue star) and the PMT. The PMT is situated a distance $ d $ from the light source and has a face with diameter $ a $ and an index of refraction that is greater than unity.}
		\label{fig:idealGeo}
	\end{figure}

A MC simulation was written to provide a proof of principle of the method discussed in the previous section. The set-up considered is shown in Figure \ref{fig:idealGeo}. A point source of light is placed centered a distance $ d $ away from the PMT face shown in light blue. The PMT face has a radius of $ a $ and an index of refraction greater than one. The MC includes a detailed model of the optics of the PMT face and uses a Lambertian distribution for the light source. 
	\FloatBarrier

\subsection{Expected Results from an Isotropic Source}
\label{sec:3.1}

The expected results from the simulation can be determined from a few simplifying assumptions. Consider an isotropic and monochromatic source of unpolarized light and the geometry depicted in Figure \ref{fig:idealGeo}. The number of photons detected is given by
	\begin{equation}
		\label{eqn:numDet0}
		N_{detected}\left( d\right) = N_{0}\epsilon\left( 1 - p_{reflect}\left( d\right) \right)\frac{\Omega\left( d\right)}{4\pi}.
	\end{equation}
Here $ N_{0} $	is the number of photons emitted by the source, $ \epsilon $ is the quantum efficiency of the detector, $ p_{reflect}(d) $ is the probability that photons from the light source are reflected off of the PMT face, and $ \Omega(d) $ is the solid angle subtended by the detector face which is given by
	\begin{equation}
		\label{eqn:solidAng}
		\Omega\left( d\right)  = 2\pi\left( 1 - \frac{d}{\sqrt{d^2+a^2}}\right).
	\end{equation}
	
While, in principle, the functional form of $ p_{reflect}\left( d\right) $ can be determined by averaging the Fresnel reflection coefficient over the incidence angles at the PMT face, to simplify the analysis tests can be performed when $ p_{reflect}\left( d\right) $ is effectively a constant. Intuitively this condition is met as the source distance becomes large since for a PMT very far away from the light source all incident photons will be approximately normal to the face. So there should be a distance at which the error incurred from assuming that $ p_{reflect}$ is constant is comparable to the other errors involved. The experimental tests will use a XP3330/B Photonis PMT which has a radius $ a=36 $\,mm and an index of refraction $ n_{detector}=1.48 $ at 420\,nm \cite{xp3330:2015}. Simulation of a detector with these parameters show that $ p_{reflect}\left( d\right) $ is approximately constant for source distances $ \geq 5$\,cm. Therefore, we can combine equations \ref{eqn:numDet0} and \ref{eqn:solidAng} to obtain
\begin{equation}
		\label{eqn:numDet1}
		N_{detected}\left( d\right) = N_{0}^{\prime}\left( 1 - \frac{d}{\sqrt{d^2+a^2}}\right).
\end{equation}
Here the constant $ N_{0}^{\prime} $ is defined as
\begin{equation}
		\label{eqn:n0Prime}
		N_{0}^{\prime} = \frac{1}{2}N_{0}\epsilon\left( 1 - p_{reflect} \right).
\end{equation}

\subsection{Simulation Proof-of-Principle}						
\label{sec:3.2}											
	The MC simulation that was written included a detailed model of the optics at the PMT's face including reflection and transmission from the photocathode as well as absorption by the photocathode. The best fit values from \cite{Motta:2004yx} were used where the XP3330/B's datasheet was incomplete. The simulation also took as its inputs the directivity of the LED \cite{R7BWD:2015}, the distance between the PMT and the source as well as angles needed to define the direction that the source points in: $ \phi_{0} $, $ \theta_{start} $, and $ \Delta\theta $, where $ \Delta\theta $ is the zenith angle step size. The code can also run with $ \phi_{0} $ sampled uniformly on $\left[ 0,2\pi\right) $. 
	
	The simulation was run with $ N_{0}=10^{6} $ photons emitted at the directions listed in Tabel \ref{tab:statWeights} for $ d =$ 5--10\,cm in 1\,cm increments. At each distance a weighted average of the number of detected photons was performed using the weights defined in Table \ref{tab:statWeights}. The results of these averages, the proxy number detected, are tabulated in Table \ref{tab:2}.
		\FloatBarrier
		\begin{table}[htbp]
			\centering
			\caption{MC results of the weighted averages of $ N_{detected} $}
			\label{tab:2}
			\begin{tabular}{ll}
			\hline \hline 
				Source Distance (cm)	&	Proxy Number Detected ($ \times10^{5} $)	\\ \hline
				5.0						&	1.57(2)										\\
				6.0						&	1.197(12)									\\
				7.0						&	0.906(9)									\\
				8.0						&	0.723(7)									\\
				9.0						&	0.593(6)									\\
				10.0					&	0.494(5)	
				\\ \hline \hline
			\end{tabular}
		\end{table}
		\FloatBarrier

\begin{figure}[h] 
		\centering
		\includegraphics[scale=0.35, angle=-90.0]{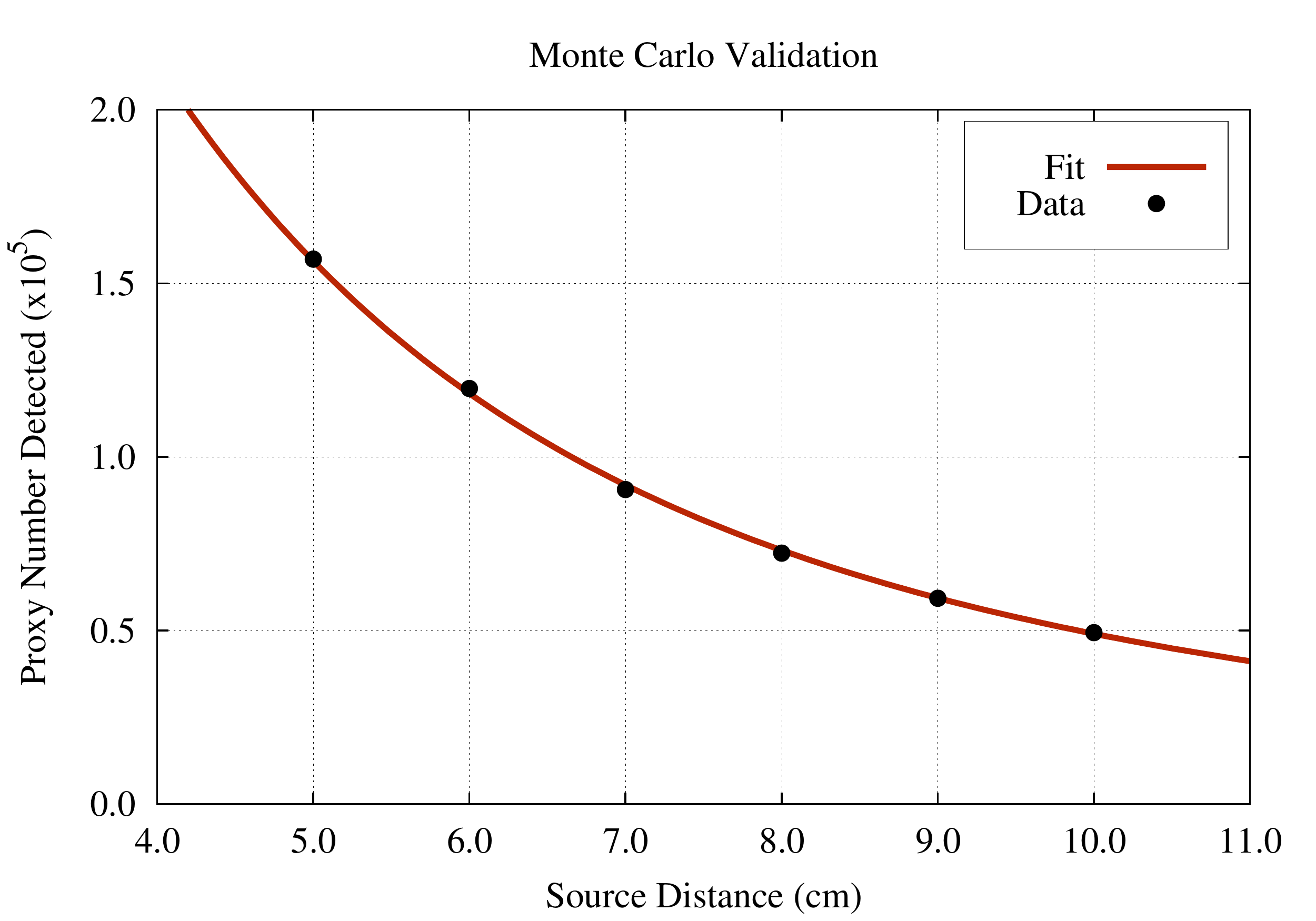} 
		\caption{Monte Carlo validation for producing an isotropic light source. Note that the errors are the same size as the plotted points. } 
		\label{fig:mcValidation}
	\end{figure}

These data can be fitted to Equation \ref{eqn:numDet1} and the goodness of the fit will determine the adequacy of the assumptions made in deriving the equation. Fitting to Equation \ref{eqn:numDet1} and holding $ a=3.6$\,cm, the best fit is $ N_{0}^{\prime} = 8.30(4)\times10^{5} $ with $ \chi^2/\mathrm{dof} = 5.5/5 $. The data and the best fit are plotted in Figure \ref{fig:mcValidation}. From the fit it is clear that the method for producing an isotropic light source by positioning an anisotropic source in different directions works as expected. 

\section{Experimental setup}
\label{sec:4_Setup}

With the method of producing an effective isotropic source from an anisotropic source validated by MC simulations, experimental validation studies were started. For these tests an R7BWD LED made by BIVAR was used. This LED has a Lambertian zenith angle emission distribution, and a peak emission wavelength of 430\,nm \cite{R7BWD:2015}. For the data runs, the directions listed above were used with $ \phi = \mathrm{constant} $ and $ \phi $ uniform on $[0,2\pi)$. 

\begin{figure}[h] 
		\centering
		\includegraphics[scale=0.4, angle=0.0]{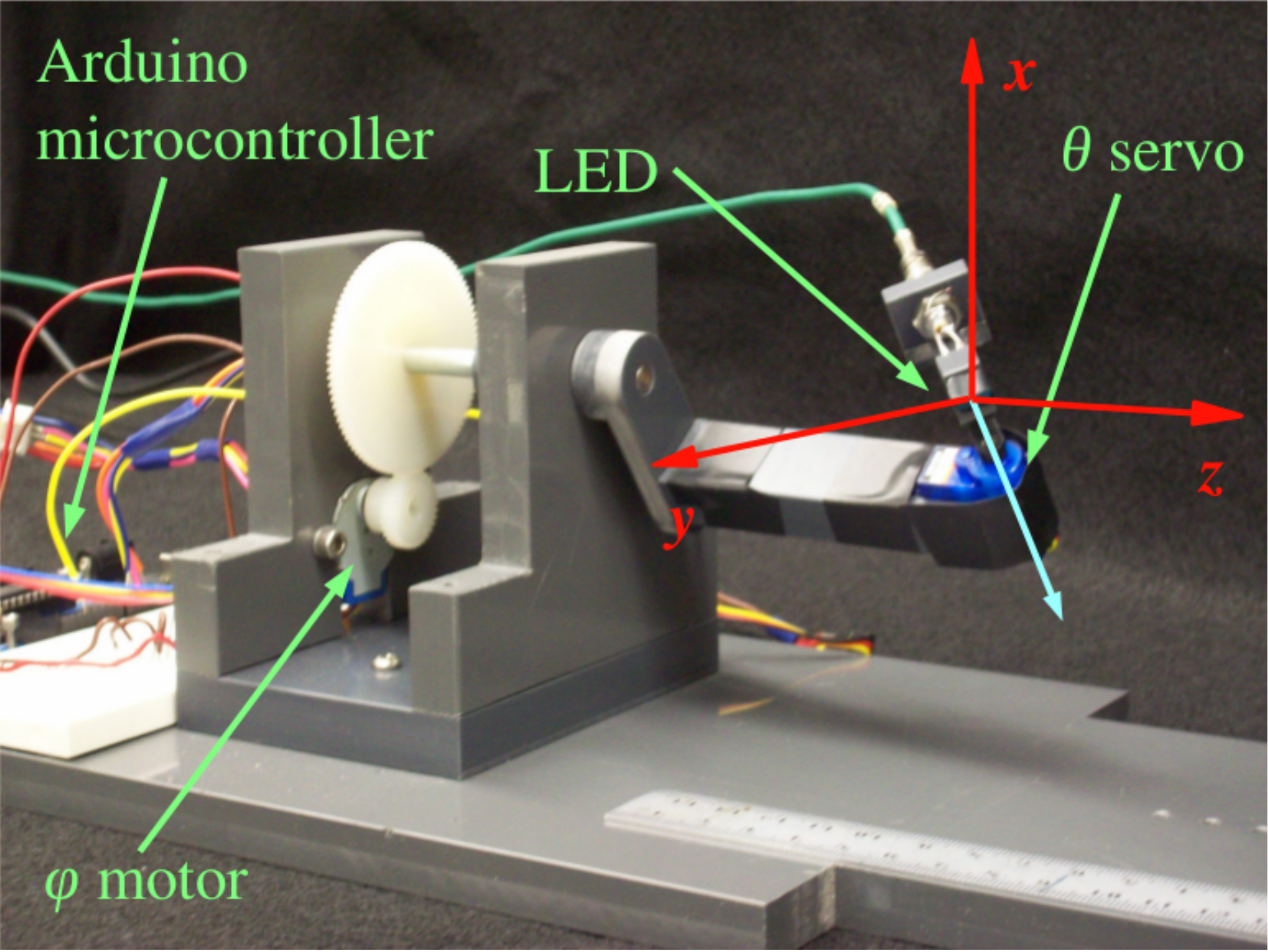} 
		\caption{Picture of the as-built light source with the various components labeled including the coordinate system assumed by the light source. Note that in the figure the light-blue arrow indicates the current direction of the LED.} 
		\label{fig:asBuilt}
	\end{figure}

The as-built light source, with its components and coordinate system, is shown in Figure \ref{fig:asBuilt}. The coordinate system's origin is centered on the front of the LED. In order to ensure that the emitted light is as close to the origin as possible, the sides of the LED were masked so that light was only emitted from the front. The LED was soldered onto a LEMO bulkhead connector which together with the LED was held in a fixture attached to the arm of a HiTec HS-55 servo; this servo sets $ \theta $. The servo was set into another fixture that is attached to a rod that was turned by a Seeedstudio RB-See-136 stepper motor which set $ \phi $.
	 
To position the LED the servo and stepper were controlled with an Arduino Uno microcontroller. The Arduino is an opensource hardware microcontroler that is operated with the use of simple programs called sketches. These sketches are written in the Arduino programming language, based on the C programing language, and uses various libraries to control attached devices \cite{Arduino}. The sketch is written in a native integrated development environment, and the compiled sketches are sent to the microcontroller via a USB cable. For the data runs, a sketch was written that positioned the servo's arm at a user inputted angle, thus setting $ \theta $ and controlled the position of the stepper motor, thus setting $ \phi $. 
	 	 
	 The LED was pulsed with fast pulses from an HP 8111OA Pulse/Pattern Generator at a rate of 300\,Hz, and the pulses at each direction were collected for 300\,s. For all the distances tested, the sizes of the pules were the same size. The size of the pulses were set such that reasonably sized pulses from the PMT could be obtained for large distances while not saturating the PMT at close distances. The pulses were run through an RG-174 cable to the LED through the bulkhead connector.
	
	\FloatBarrier
	\FloatBarrier	
	
	 To understand the uncertainties related to the reproducibility of the data, runs at every distance and direction were taken at least twice. Finally, the waveforms from the PMT were recorded with a CAEN v1721 8-bit digitizer using in-house DAQ software. The digitizer was triggered with a TTL pulse produced simultaneously with the LED pulse.

\section{Results}
\label{sec:5_Results}

	 The prescription outlined in Section \ref{sec:2_Concept} was used to test the as-built light source. Data was taken for each direction and for the PMT at distances of 5--10\,cm in 1\,cm increments with $ \phi $ constant and with $ \phi $ uniform on $ [0,2\pi) $. Once the data was taken, the pulse integrals were determined by summing the samples of the waveform and subtracting the integral of the baseline. The baseline was determined by setting the acquisition window to be much larger than the pulse duration and taking the average of the first and last 100 samples of each event. Since the pulses driving the LED were of the same size (in both voltage and timing) the same amount of light was generated by the LED for each pulse. Therefore, for a given source distance and source direction, the histograms of the integrals have an approximately Gaussian shapes. 
	 	 
	 Once the integrals were determined, the integrals for each direction and source distance were binned in histograms. Next, cuts where chosen to eliminate dark-current and other non-signal pulses to produce a histogram with the signal peak only. These were fit with a Gaussian and the best fit's centroid was used in further analysis. Repeated runs were then averaged, and, as a measure of the centroid's uncertainty, the sample standard deviations of these were calculated. With the centroids determined for each distance and direction, the weighted average of directions discussed above was performed for each distance. The results, the proxy integrals, in units of V\,ns, are listed in Table \ref{tab:dataResults}.
	 
\begin{figure}[h] 
		\centering
		\includegraphics[scale=0.35, angle=-90.0]{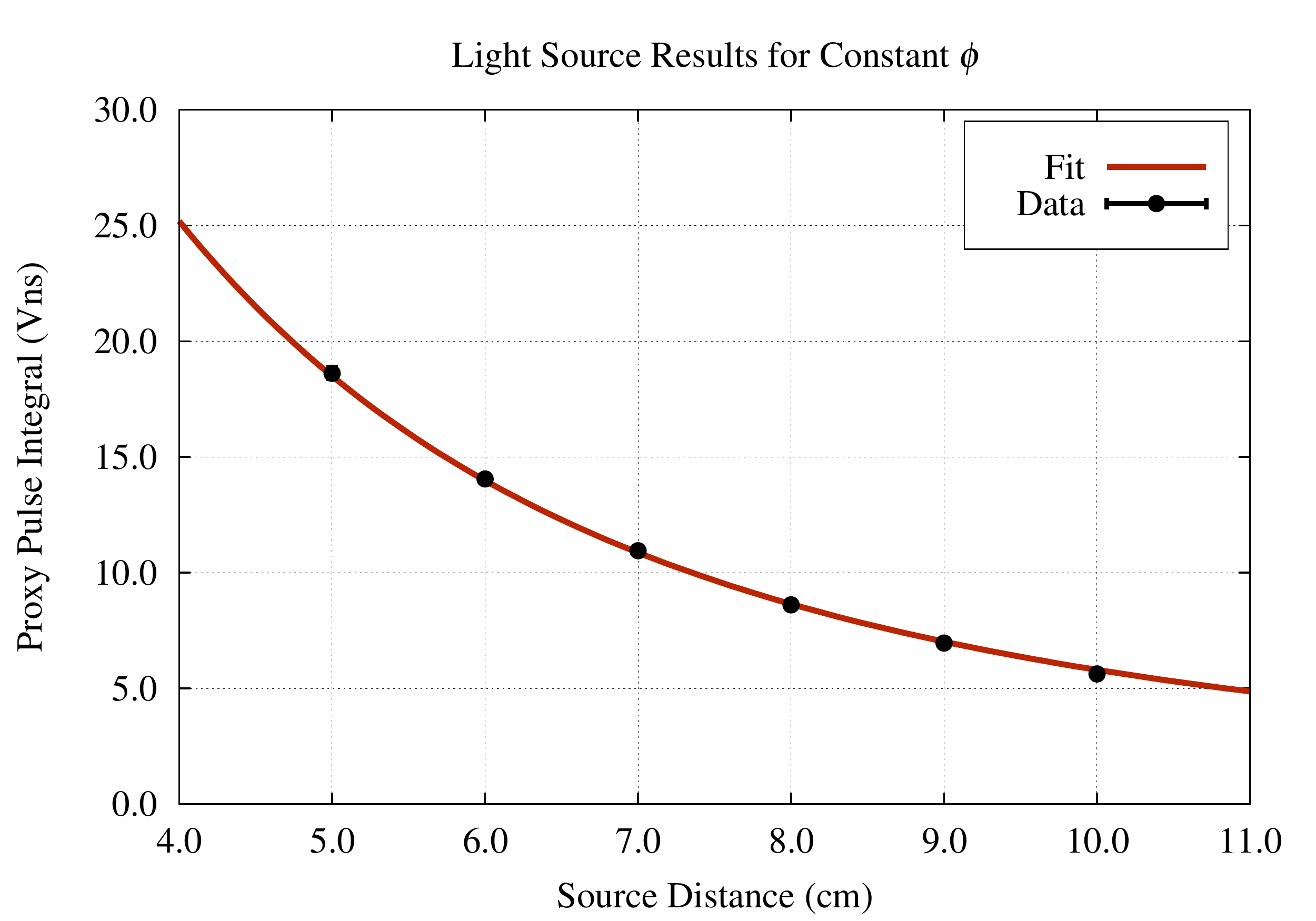} 
		\caption{The results for constant $ \phi $. Note that the error bars are the same size as the plotted points.}
		\label{fig:dataResultsConst}
	\end{figure}

\begin{figure}[h!] 
		\centering
		\includegraphics[scale=0.35, angle=-90.0]{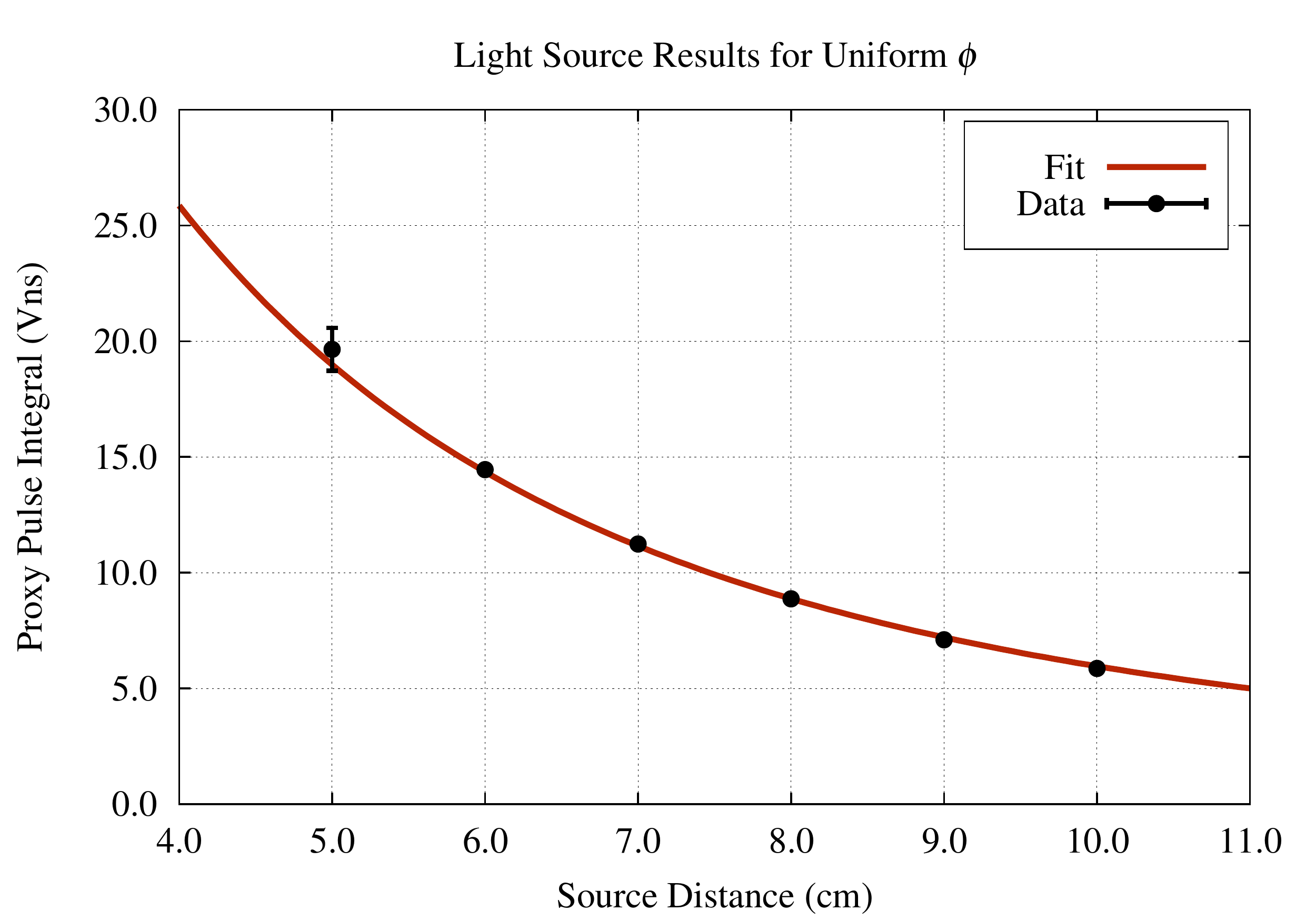} 
		\caption{The results for uniform $ \phi $. Except for the point at 5\,cm the error bars are of the same size as the plotted points.}
		\label{fig:dataResultsUni}
	\end{figure}

	 	\FloatBarrier
		\begin{table}[h]
			\centering
			\caption{Proxy integrals for initial tests}
			\label{tab:dataResults}
			\begin{tabular}{lll}
				\hline \hline
				Source Distance (cm)	&	$ \phi $ Constant (V\,ns)	&  $ \phi$ Uniform (V\,ns) \\ \hline
				5.0		&	18.6(2)		&	19.6(9)		\\
				6.0		&	14.05(13)	&	14.4(2)		\\
				7.0		&	10.94(12)	&	11.24(16)	\\
				8.0		&	8.60(8)	&	8.87(15)	    \\
				9.0		&	6.96(12)	&	7.10(14)	\\
				10.0	&	5.62(12)	&	5.86(10)
				\\ \hline \hline
			\end{tabular}
		\end{table}
		\FloatBarrier	 
The data in Table \ref{tab:dataResults} can be fitted with the expected curve for an isotropic source. The data and their fits are shown in Figures \ref{fig:dataResultsConst} and \ref{fig:dataResultsUni}. The fits give $ N_{0}^{\prime}=98.1(6) $ and $ N_{0}^{\prime}=101(2) $ for uniform and constant $ \phi $ respectively; the $ \chi^2/\mathrm{dof} $'s are $2.6/5$ and $3.9/5$ indicating a good fit to an isotropic light source. This is also clear from the figures. Note, however, that while one expects the fits for both cases to be equivalent they differ by greater than $ 1\sigma $. This discrepancy is not that significant and probably due to errors associated with reproducing the setup between runs, or to an unknown systematic error caused by picking a specific $ \phi $. Future work can check for any systematic dependence on the choice of $ \phi $.

    \FloatBarrier

%

\section{Conclusions}
\label{sec:6_Conclusions}

	We have demonstrated, via MC simulation and an as-built system, the production of a light source that effectively behaves as if it were isotropic by positioning an LED in several directions and computing the weighted average of the detector responses. Using this well-characterized source the integral performance of optical components, like PMTs and light guides, can be accurately measured on an individual basis. While the success of this particular method for producing an isotropic light source is an important new tool in nuclear instrumentation, there are, in principle, many ways to achieve the same effect. Additionally, a similar method can be used to approximate any desired emission distribution.  Therefore, this method can be optimized the particular application, and the future applications of this method may prove useful for the community of scientists using optical nuclear instrumentation. 
	
\section*{Acknowledgements}
We would like to thank Tristan Wright for his help and electronics support for this project and the Virginia Tech Physics Department machine shop for their work with the design and machining of the robotic light source. This work was funded by the National Science Foundation award number 1001394.

\section*{References}

\bibliography{references}{}

\end{document}